\documentclass[letterpaper,twocolumn,10pt]{usetex-v1}

\usepackage{times}
\usepackage{graphicx}
\usepackage{algorithm}
\usepackage{algpseudocode}
\usepackage{color}
\usepackage{endnotes}
\usepackage{listings}
\usepackage{makecell}
\usepackage{multirow}
\usepackage[hyphens]{url}
\usepackage{cite}
\usepackage[colorlinks=false,linkcolor=black,citecolor=black,pdfborder={0 0 0}]{hyperref}
\usepackage{breakurl}
\usepackage{enumitem}
\usepackage{xcolor}
\usepackage{xspace}
\usepackage[kerning,spacing]{microtype}
\usepackage[font=small,labelfont=bf]{caption}
\usepackage{amsmath}
\usepackage[scaled=0.92]{helvet}

\usepackage{mathptmx}
\DeclareMathAlphabet{\mathcal}{OMS}{cmsy}{m}{n}
\DeclareMathAlphabet{\mathrm}{OT1}{bch}{m}{n}
\DeclareMathAlphabet{\mathit}{OT1}{bch}{m}{it}

\setitemize{noitemsep,topsep=3pt,parsep=0em,partopsep=3pt}

\newcommand{\sysname}{HHZS\xspace}
\renewcommand{\paragraph}[1]{{\smallskip\noindent\bf #1}}

\DeclareMathOperator*{\argmin}{arg\,min}

\begin{document}

\title{Efficient LSM-Tree Key-Value Data Management on \\
	Hybrid SSD/HDD Zoned Storage}
\author{Jinhong Li, Qiuping Wang, Patrick P. C. Lee \\ {\em The Chinese University of Hong Kong}}

\maketitle

\begin{abstract} 
Zoned storage devices, such as zoned namespace (ZNS) solid-state drives (SSDs)
and host-managed shingled magnetic recording (HM-SMR) hard-disk drives (HDDs),
expose interfaces for host-level applications to support fine-grained,
high-performance storage management. Combining ZNS SSDs and HM-SMR HDDs into a
unified hybrid storage system is a natural direction to scale zoned storage at
low cost, yet how to effectively incorporate zoned storage awareness into
hybrid storage is a non-trivial issue.  We make a case for key-value (KV)
stores based on log-structured merge trees (LSM-trees) as host-level
applications, and present \sysname, a middleware system that bridges an
LSM-tree KV store with hybrid zoned storage devices based on {\em hints}.
\sysname leverages hints issued by the flushing, compaction, and caching
operations of the LSM-tree KV store to manage KV objects in placement,
migration, and caching in hybrid ZNS SSD and HM-SMR HDD zoned storage.
Experiments show that our \sysname prototype, when running on real ZNS SSD and
HM-SMR HDD devices, achieves the highest throughput compared with all baselines
under various settings. 
\end{abstract}

\section{Introduction}
\label{sec:intro}

Conventional storage software stacks leverage the block interface to bridge
host-level applications and storage devices, but the block interface poses
performance penalties due to its mismatch with modern storage hardware
characteristics.  Specifically, both NAND-flash-based solid-state drives (SSDs)
and Shingled Magnetic Recording (SMR) hard-disk drives (HDDs) build on
append-only writes by nature. To comply with the block interface, they must be
coupled with a translation layer that maps the logical addresses of application
data to the physical addresses of storage devices.  Such a translation layer
also necessitates device-level garbage collection, which incurs I/O
interference and performance variability to applications.  {\em Zoned storage}
\cite{zonedstorage,han21,aghayev20,bjorling21} has been advocated to replace
the block interface with a new zoned interface.  By exposing the zoned
interface to applications and letting applications have fine-grained control in
storage management, zoned storage not only maintains performance predictability
by freeing the costly translation layers and hence on-device garbage collection
\cite{bjorling21}, but also enables applications to customize the software
stacks and exploit the full performance potential of modern storage hardware.

Zoned namespace (ZNS) SSDs and host-managed SMR (HM-SMR) HDDs are two
mainstream zoned storage devices available today. ZNS SSDs offer better I/O
performance, while HM-SMR HDDs can provide high capacity at much lower cost
({\S\ref{subsec:motivation}).  It is natural to support {\em hybrid} zoned
storage to combine the benefits of both types of devices, while preserving
performance predictability by eliminating translation layers via zoned storage. 
	

Hybrid storage is a well-studied topic in the literature (e.g.,
\cite{li19ursa,xie18,xie19,chen21}), yet existing hybrid storage solutions
still build the block interface for storage management.  While some wisdom in
conventional hybrid storage can be applied to hybrid zoned storage (e.g., 
storing frequently accessed data in high-performance devices), there are
unique design challenges in hybrid zoned storage.  In particular, zoned storage
devices organize data in units of {\em zones} of hundreds of MiB, while the
data in a zone must be {\em reset} at once before being overwritten.  If the
data objects of different lifetimes are stored in the same zone, there will be
both high space amplification due to the occupied space by stale data and high
write amplification due to the relocation of live data from a reset zone.
Hybrid zoned storage incurs additional data movement between heterogeneous
zoned storage devices (i.e., ZNS SSDs and HM-SMR HDDs).  An effective hybrid
zoned storage design should incorporate zone awareness into the data
management of hybrid zoned storage devices.  


Our insight is that applications now have fine-grained control in zoned
storage management as opposed to legacy storage hardware, so they can provide
{\em hints} \cite{patterson95} (e.g., application-level semantics and access
patterns) for instructing how zoned storage devices should manage data.
Application hints have been extensively used to improve storage performance,
such as data prefetching and caching \cite{patterson95,liu09,kumar20}, CPU
consumption reduction \cite{tai21}, and throughput improvements
\cite{sarkar96,ouyang14}.  In this work, we make a case for key-value (KV)
storage based on log-structured merge trees (LSM-trees), and show how hints
facilitate the deployment of LSM-tree KV stores atop hybrid zoned storage.
LSM-tree KV stores are good fits for zoned storage since they issue
append-only writes to physical storage (in KV objects)
\cite{yao19,bjorling21}.  In fact, RocksDB \cite{rocksdb}, a production
LSM-tree KV store, exports the write lifetime hint for zone selection
\cite{bjorling21}.  Nevertheless, the usage of hints in RocksDB remains
preliminary, not to mention hybrid zoned storage. 

In this paper, we present \sysname, a middleware system that implements hinted
hybrid zoned storage that bridges the upper-layer LSM-tree KV store with the
underlying hybrid ZNS-SSD and HM-SMR HDD zoned storage.  \sysname leverages
hints provided by the internal operations (e.g., flushing, compaction, and
caching) of the LSM-tree KV store (\S\ref{subsec:lsm}) to address three 
data management aspects in hybrid zoned storage: (i) data placement in
different zoned storage devices on the write path, (ii) data migration across
zoned storage devices in the background, and (iii) caching of frequently
accessed data in SSD storage. 

We prototyped \sysname by modifying the LSM-tree KV store RocksDB
\cite{rocksdb} and the zone-aware file system ZenFS \cite{zenfs}.  We evaluate
our \sysname prototype on real ZNS-SSD and HM-SMR HDD devices, and show that
\sysname achieves the highest throughput compared with the baselines under
various workloads and parameter settings.  For example, under the six YCSB core
workloads \cite{cooper10}, \sysname achieves 28.0-69.3\% higher throughput than
the automated placement scheme in SpanDB (an LSM-tree KV store for hybrid
storage) \cite{chen21}.  

\section{Background}
\label{sec:background}

We provide the background details on zoned storage (\S\ref{subsec:zoned}) and
LSM-trees (\S\ref{subsec:lsm}).  We also present analysis to show the
limitations of basic data placement schemes (\S\ref{subsec:motivation}) and 
motivate the challenges of deploying LSM-tree KV stores on hybrid zoned
storage (\S\ref{subsec:challenge}). 

\subsection{Zoned Storage}
\label{subsec:zoned}

Zoned storage offers higher storage capacities, increased throughput, and
lower latencies via the cooperation of host-level applications and storage
devices \cite{zonedstorage}.  There are two main types of zoned storage
devices, ZNS SSDs and HM-SMR HDDs.  ZNS SSDs build on NAND-flash-based SSDs, in
which each flash page must be erased before being programmed with new written
data.  As erase operations must be performed in units of blocks with hundreds
of flash pages each, ZNS SSDs prohibit in-place updates and issue writes in an
append-only manner.  HM-SMR HDDs build on the SMR technologies that overlap
adjacent disk tracks for substantially increasing the disk areal density.
In-place updates can disturb the data in adjacent disk tracks. Thus, HM-SMR
HDDs also issue writes in an append-only manner.  As a result, zoned storage
respects the physical natures of ZNS SSDs and HM-SMR HDDs by exposing
append-only writes to applications.  Recent studies show that applications with
zoned storage interfaces achieve better performance than with block interfaces
(e.g., 44\% higher throughput and 50\% less tail latency using a ZNS SSD than
using a multi-stream SSD \cite{bjorling21}, and 2.38$\times$ throughput gain
using an HM-SMR HDD compared with a conventional HDD \cite{yao19}).  

Zoned storage unifies the abstraction of both ZNS SSDs and HM-SMR HDDs.  It
divides the physical address space into append-only {\em zones}, each of which
has a writable {\em zone capacity} of hundreds of MiB (e.g., 1,077\,MiB for ZNS
SSDs \cite{bjorling21} and 256\,MiB for HM-SMR HDDs \cite{yao19}).  Each
zone represents a contiguous address space. It can be read in any order, but
can only be written sequentially.  It is associated with a {\em write
pointer}, which indicates the offset of the next write. The write
pointer initially references the beginning of a zone, and moves forward by the
number of bytes written for each write request.  Zoned storage exposes zones
to applications, which can then specify the zones via which data is read or
written.  To update (or overwrite) any data in a zone, applications must first
issue a {\em reset} command to make the write pointer reference again the
beginning of the zone. 

In this work, we explore hybrid zoned storage that comprises both ZNS SSDs and
HM-SMR HDDs. For brevity, our discussion refers to ZNS SSDs and HM-SMR HDDs as
SSDs and HDDs, respectively if the context is clear.


\subsection{LSM-Trees}
\label{subsec:lsm}

LSM-tree KV stores issue the writes or updates of KV objects as sequential
writes to storage.  They are friendly to zoned storage that builds on
append-only writes \cite{yao19,bjorling21}.  We now provide a high-level
overview of a typical LSM-tree KV store (e.g., LevelDB \cite{leveldb} and
RocksDB \cite{rocksdb}), as shown in Figure~\ref{fig:lsm}. 

\begin{figure}[!t] 
\centering
\includegraphics[width=3.2in]{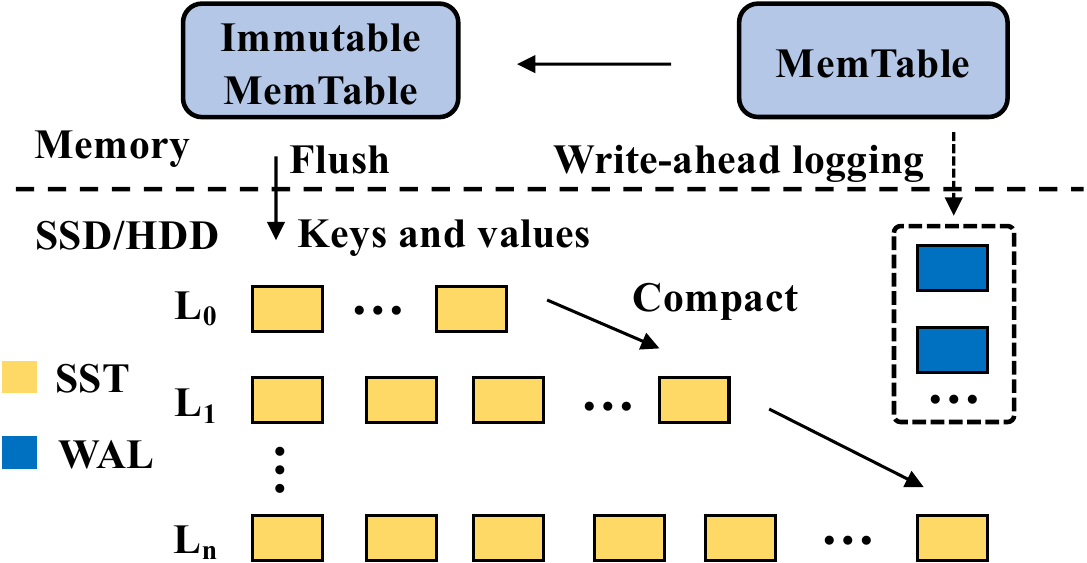} 
\vspace{-3pt}
\caption{Architecture of a typical LSM-tree KV store.}
\label{fig:lsm}
\vspace{-6pt}
\end{figure}

\paragraph{LSM-tree structure.} An LSM-tree (Figure~\ref{fig:lsm}) organizes
KV objects in $n+1$ levels, referred to as $L_0, L_1, \cdots, L_n$, where
$L_j$ is a higher level than $L_i$ for $0\le i < j \le n$.  It maintains KV
objects in immutable files called {\em SSTables (SSTs)} of several MiB each
(e.g., 64\,MiB in RocksDB).  Each SST is divided into multiple {\em data
blocks} (of several KiB each) and maintains an {\em index block} that stores
the key ranges and the offsets of all data blocks in the SST.  To support
efficient range queries, each SST keeps all KV objects sorted by keys, and all
SSTs at the same level (except $L_0$) have disjoint key ranges.  

Each level is configured with a {\em target size} that specifies the expected
maximum amount of KV objects being stored.  The target sizes across levels
typically increase exponentially; for example, the default target size of
$L_{i+1}$ is $10\times$ that of $L_i$ for $i\ge 1$ in LevelDB \cite{leveldb}
and RocksDB \cite{rocksdb}.  

\paragraph{Writes.}  To write a KV object, an LSM-tree KV store appends the KV
object into both an on-disk {\em write-ahead log (WAL)} for crash consistency
and an in-memory write buffer called the {\em MemTable} for batched writes
(the LSM-tree KV store may allocate multiple MemTables).
When the size of a MemTable reaches a limit
(e.g., 64\,MiB by default in RocksDB \cite{rocksdb}), the LSM-tree KV store
makes the MemTable immutable,  flushes the immutable MemTable as an SST
into $L_0$, and deletes the flushed KV objects from the WAL.  If the size of a
level $L_i$ ($i\ge 0$) exceeds its target size, the LSM-tree KV store triggers
{\em compaction} to merge the SSTs of $L_i$ and the next higher level
$L_{i+1}$.  Specifically, it retrieves an SST from $L_i$ and the SSTs at
$L_{i+1}$ that have overlapping key ranges with the SST from $L_i$. It then
merge-sorts the KV objects in the retrieved SSTs, and discards any older
versions of a KV object.  Finally, it writes back all sorted KV objects as new
SSTs into $L_{i+1}$.  Both flushing and compaction operations are done in the
background, and multiple compaction operations can be issued simultaneously if
they do not have SSTs with overlapping key ranges.  

\paragraph{Reads.} To read a KV object, an LSM-tree KV store first searches
for the key in all MemTables.  If unsuccessful, it searches each level,
starting from $L_0$, until the key is found at a level or does not exist at
all levels.  Since each level is sorted, searching for a KV object at a level
can be done by binary search: (i) it first searches for the SST whose key
range covers the searched key, (ii) it then searches the index block for the
data block that possibly contains the searched key, and (iii) it finally
searches the KV object within the data block.  To accelerate reads, each SST
maintains a Bloom filter \cite{bloom70} and the search within an SST is done
only if the key probably exists.  The frequently accessed data blocks and
index blocks can also be cached in an in-memory {\em block cache} (e.g., in
RocksDB) to mitigate the disk I/O overhead.

\subsection{Motivating Analysis}
\label{subsec:motivation}

\begin{table}[t]
\centering
\small
\begin{tabular}{c|c|c}
\hline
	   & \makecell{ZN540 \\ (ZNS SSD)} 
           & \makecell{ST14000NM0007 \\ (HM-SMR HDD)} \\
\hline
\hline
{\bf Sequential reads (MiB/s)}   & 1039.6     & 210.0 
\\ \hline
{\bf Sequential writes (MiB/s)}   & 1002.8    & 210.0 
\\ \hline
{\bf Random reads (IO/s)}   & 16928.3  & 115.0 
\\ \hline
{\bf Price (US\$/GiB)}          & 
    0.28  \cite{zn540,sn640} & 
    0.021 \cite{seagate_hmsmr}
\\ \hline
\end{tabular}
\caption{Performance statistics and prices in May 2022 of our ZNS SSD and
HM-SMR HDD devices (the price of the ZNS SSD is estimated with that of an NVMe
SSD).} 
\label{tab:perf}
\vspace{-6pt}
\end{table}

\begin{figure*}[t]
\begin{tabular}{@{\ }ccc}
\includegraphics[width=2.2in]{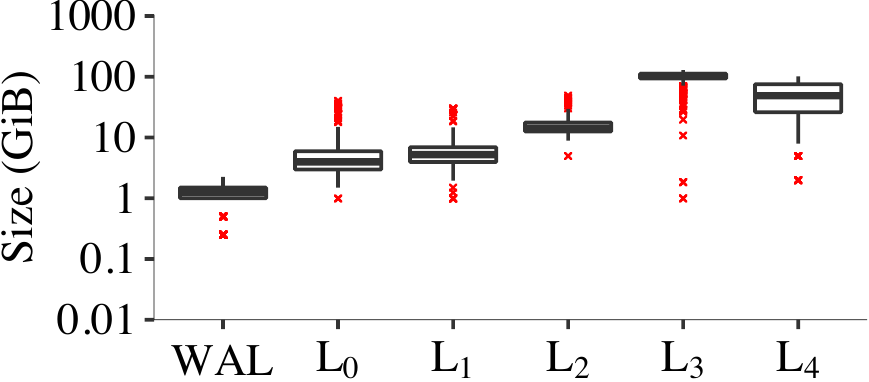} &
\includegraphics[width=2.2in]{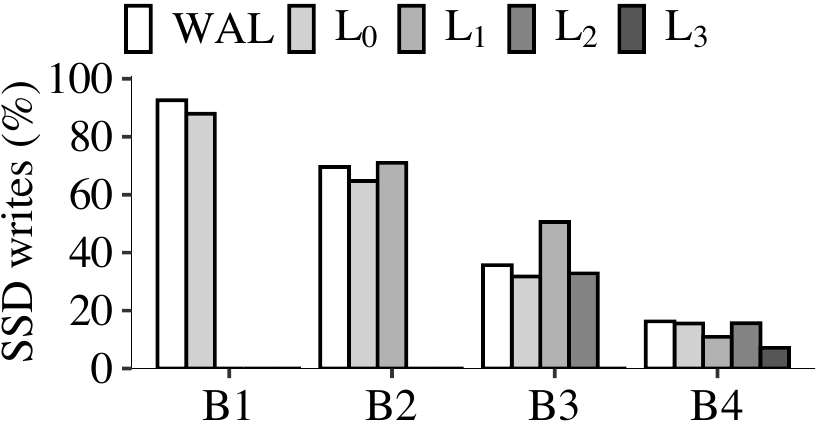} & 
\includegraphics[width=2.2in]{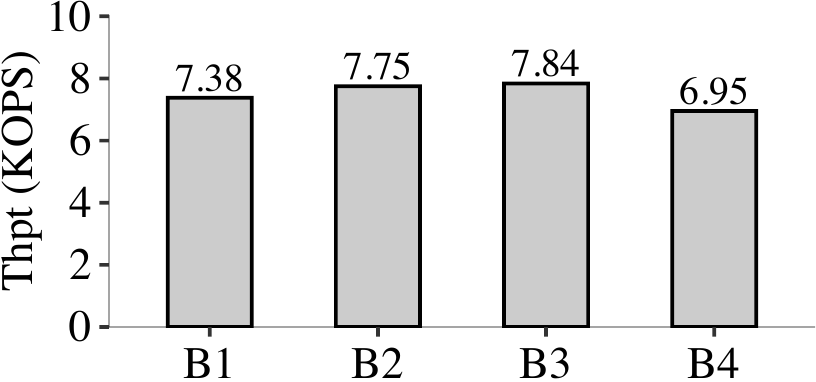} \\
\mbox{\small (a) Actual data sizes in B4}  & 
\mbox{\small (b) Percentages of SSD write traffic}  & 
\mbox{\small (c) Load throughput} 
\vspace{3pt}\\
\includegraphics[width=2.2in]{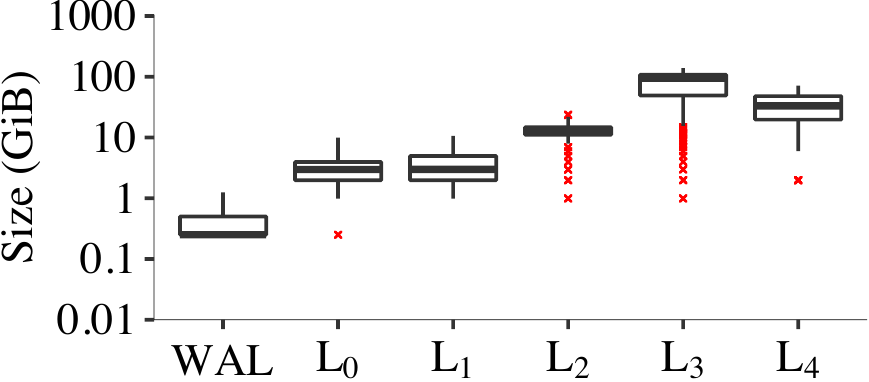} &
\includegraphics[width=2.2in]{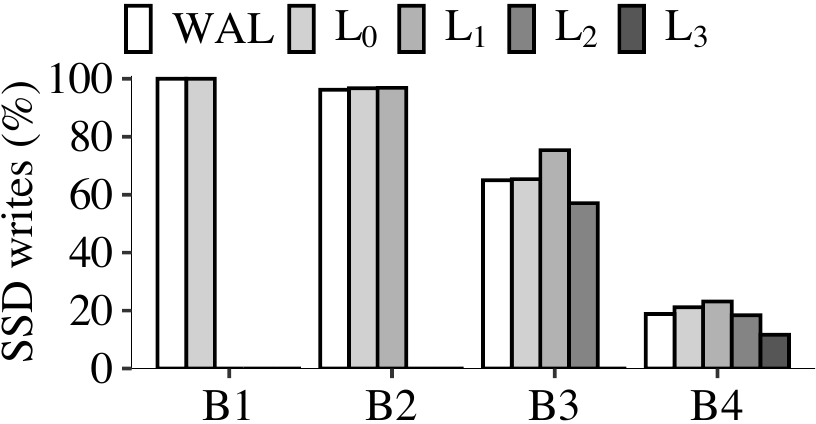} & 
\includegraphics[width=2.2in]{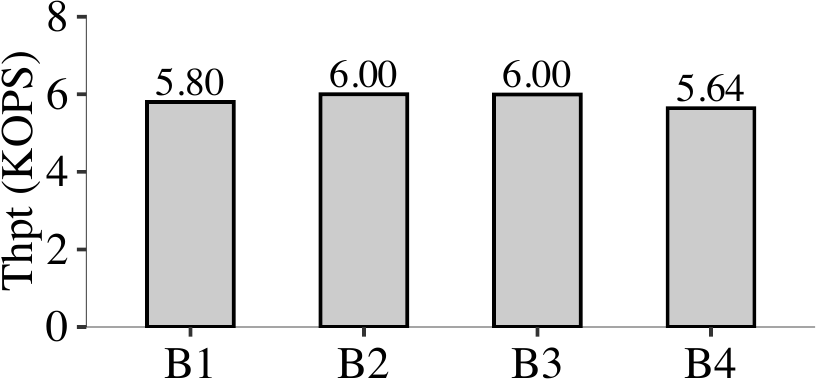} \\
\mbox{\small (d) Actual data sizes in B4 (throttled)}  & 
\mbox{\small (e) Percentages of SSD write traffic (throttled)}  & 
\mbox{\small (f) Load throughput (throttled)} 
\vspace{3pt}\\
\includegraphics[width=2.2in]{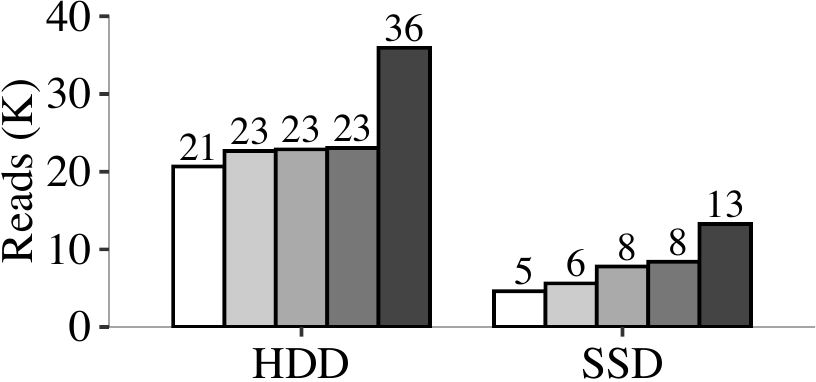} & 
\includegraphics[width=2.2in]{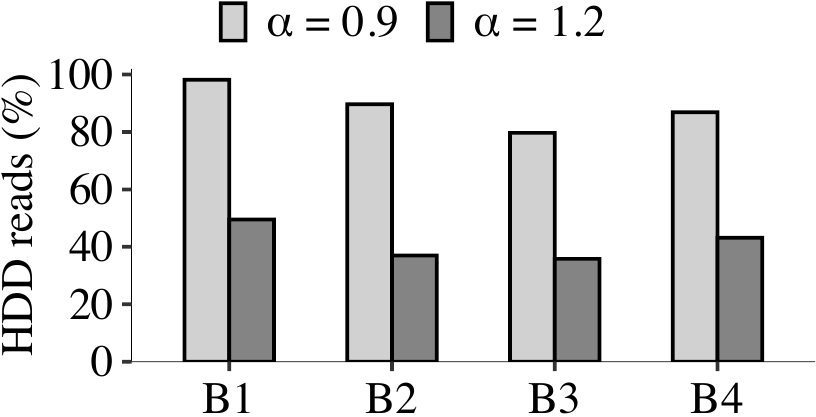} & 
\includegraphics[width=2.2in]{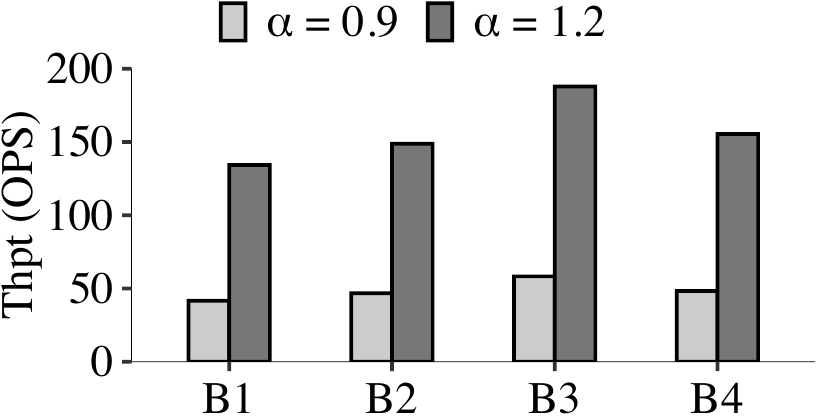} \\
{\small (g) Top-five SST reads at $L_3$ in B4} &
{\small (h) Percentages of HDD read traffic} &
{\small (i) Read throughput} 
\end{tabular}
\caption{Measurement on basic data placement schemes: first loading 200\,GiB
of 1-KiB KV objects without throttling (figures~(a)-(c)) and with throttling
(figures~(d)-(f)), followed by 1\,M reads with skewness
(figures~(g)-(i)).}
\label{fig:motivation}
\end{figure*}

Hybrid zoned storage requires effective data management across zoned storage
devices. We now motivate via experiments that na\"ive data management can lead
to significant performance degradations.  

We consider two real zoned storage devices, a 4-TiB Western Digital Ultrastar
DC ZN540 ZNS SSD \cite{zn540} and a 14-TiB Seagate ST14000NM0007 HM-SMR HDD
\cite{seagate_hmsmr}, with the zone capacities of 1,077\,MiB and
256\,MiB, respectively.  We analyze the cost-performance trade-off of both
devices, as shown in Table~\ref{tab:perf}.  We measure the throughput of
sequential reads/writes (with 1\,MiB requests) and random reads (with 4\,KiB
requests) for both devices, using \texttt{fio} \cite{fio} with a queue
depth of one.  We also report the prices (in US\$/GiB) for both devices;
since the price of the ZNS SSD is not yet publicly available at the time of
the writing, we use the 3.84-TiB Ultrastar DC SN640 block-interface NVMe SSD
for cost estimation given that both ZN540 and SN640 models have similar
throughput and the same form factor \cite{zn540, sn640}.  In short, the
ZNS SSD provides much higher throughput than the HM-SMR HDD (e.g.,
147.2$\times$ higher in random reads), yet its price per GiB is 13.1$\times$
higher.  This motivates us to explore hybrid zoned storage to balance the
cost-performance trade-off. 

We set up a hybrid zoned storage environment as follows.  We mount RocksDB
\cite{rocksdb} on ZenFS \cite{zenfs}, which we have modified to support hybrid
zoned storage devices (\S\ref{subsec:impl}). We configure RocksDB based on the
RocksDB tuning guide \cite{rocksdb_guide}, and run the modified ZenFS on the
both devices.  

We further implement a set of {\em basic} data placement schemes to understand
their limitations.  The basic schemes make the placement decision based on (i)
the filename of the written file and (ii) the LSM-tree level of the SST (if the
written file is an SST).  We use the filename, which is specified by RocksDB
during file writes, to determine if the written file belongs to the WAL or 
a newly written SST.  For the latter case, we modify the {\tt FileOptions}
structure in RocksDB to include the LSM-tree level of the newly written SST. 
Then the basic schemes attempt to store both the WAL and the low-level SSTs in
the SSD, and the high-level SSTs in the HDD.  The rationale is that the WAL
lies on the critical I/O path and WAL writes are perceived by users, so the
WAL should be stored in high-speed storage \cite{chen21}.  Also, the low-level
SSTs store the recently written KV objects that are more likely accessed than
the high-level SSTs, so they are also stored in the SSD for low-latency
access.  Note that the similar rationale is also adopted in \cite{chen21}.  To
avoid extra I/Os of data migration and hence long stalls, if the SSD is full,
the basic schemes simply issue the writes of the WAL or low-level SSTs to the
HDD, and the SSD space can be reused if some SSTs are removed by the LSM-tree
compaction. 

In our evaluation, we vary the level threshold, denoted by $h$, that
differentiates the low and high LSM-tree levels.  Let B$h$ be the basic scheme
that puts the SSTs at levels $L_0, \cdots, L_{h-1}$ in the SSD and the SSTs at
levels $L_h$ or higher in the HDD.  We make four key observations on the
basic schemes. 


\paragraph{O1:} {\em The actual size of SSTs at each level can significantly
exceed the target size during runtime.} 

We first examine whether the basic schemes (where $1\le h\le 4$) can
effectively put the low-level SSTs in the SSD.  We use YCSB \cite{cooper10}
to load into RocksDB 200\,GiB of 1-KiB KV objects, each with a 24-byte key
and a 1,000-byte value.  We limit the available SSD space as 20 zones
(i.e., 20 $\times$ 1,077\,MiB $=$ 21.0\,GiB) and set the target sizes of
$L_0$, $L_1$, $L_2$, $L_3$, and $L_4$ as 1\,GiB, 1\,GiB, 10\,GiB, 100\,GiB,
and 1,000\,GiB, respectively.

We sample the actual sizes of the WAL and $L_0-L_4$ at one-minute intervals
when the KV objects are loaded (which takes about eight hours in total).
Figure~\ref{fig:motivation}(a) shows the boxplots (including the minimum, lower
quartile, median, upper quartile, and maximum) of the samples from B4; the
results of B1-B3 are similar and omitted from our plots. The maximum sizes of
$L_0$, $L_1$, and $L_2$ can reach 40.3\,GiB, 29.7\,GiB, and 49.1\,GiB, or
equivalently $40.3\times$, $29.7\times$, and $4.9\times$ their target sizes,
respectively, and their medians also achieve $4.0\times$, $5.2\times$, and
$1.4\times$ their target sizes, respectively. The reason that the actual size
can grow beyond the target size is that the compaction operations cannot catch
up with the write speed under write-intensive workloads due to the limited I/O
bandwidth of the SSD/HDD.  Thus, it is ineffective to statically determine the
reserved space for each level based on its target size.  

\paragraph{O2: } {\em The load throughput degrades when $h$ is too small since 
the SSD is not fully utilized, or when $h$ is too large since the WAL and
low-level SSTs cannot be entirely stored in the SSD.}   

We study the throughput (in operations per seconds (OPS)) of loading KV
objects by tuning $h$.  
Figure~\ref{fig:motivation}(b) shows the percentage of write traffic that
goes to the SSD with respect to the total write traffic for each basic scheme
B$h$; recall that B$h$ may issue the writes of the WAL and low-level SSTs
(i.e., at levels from $L_0$ to $L_{h-1}$) to the HDD if the SSD is full.  For
$h=1$ or $h=2$, both B1 and B2 can issue most of the writes of the WAL and
low-level SSTs to the SSD.  B2 has higher throughput than B1 as it can put the
SSTs at $L_1$ in the SSD as well (but not for B1).  However, for $h=4$, large
fractions of the SSTs at $L_0$ and $L_1$ are written to the HDD (84.4\% and
89.1\%, respectively).  The reason is that some SSTs at level $L_3$ occupy the
SSD space and prohibit the future writes of SSTs at $L_0$ and $L_1$ to the
SSD. 

Figure~\ref{fig:motivation}(c) shows the load throughput for the basic
schemes.  B3 has the highest load throughput among all basic schemes, while B4
has 11.3\% lower throughput than B3 even though it attempts to put
more low-level SSTs in the SSD.  However, B3 still has significant fractions of
writes of the WAL (68.2\%) and low-level SSTs (68.2\% at $L_0$, 49.4\% at
$L_1$, and 67.2\% at $L_2$) that go to the HDD.

\paragraph{O3: } {\em Throttling writes cannot address the limitations of the
basic data placement schemes.}

To prevent writes from outpacing compaction, one possible solution is to
rate-throttle writes.  Here, we use the {\tt -target} option in YCSB to
throttle the rate of writing KV objects to 6,000\,OPS, which is smaller than
the load throughput of the basic schemes without throttling as shown in 
Figure~\ref{fig:motivation}(c).
Figure~\ref{fig:motivation}(d) shows the boxplots of the samples from B4 with
the throttled loading rate of 6,000\,OPS.  Unfortunately, the actual sizes of
$L_0$, $L_1$, and $L_2$ still exceed their target sizes, with the maximum
sizes of $10.0\times$, $10.7\times$, and $2.4\times$ their target sizes,
respectively.  While the numbers are smaller than without throttling
(Figure~\ref{fig:motivation}(a)), throttling writes not only degrades
performance, but also cannot entirely prevent an LSM-tree level from having a
larger actual size than its target size. 

Figure~\ref{fig:motivation}(e) shows the percentage of write traffic to
the SSD, and Figure~\ref{fig:motivation}(f) shows the load throughput, when
writes are throttled to 6,000\,OPS.  We find similar observations to without
throttling.  In particular, for B3, even with write throttling, the total
sizes of WAL and SSTs in $L_0$, $L_1$, and $L_2$ can exceed the total SSD
size, so some WAL data or $L_0$ files will be stored in the HDD
(Figure~\ref{fig:motivation}(e)).  Also, we find similar throughput
degradations for B1 and B4 compared with B2 and B3 (whose throughput is
upper-bounded at 6,000\,OPS due to throttling), since too few SSTs (in B1)
or too many SSTs (in B4) are written to the SSD. 

\paragraph{O4: } {\em  Frequently read SSTs can reside in the HDD, thereby
leading to the degraded read throughput.}

Recall that the basic schemes may put low-level SSTs in the HDD if the SSD
becomes full.  We show how this phenomenon degrades read performance.  After
loading 200\,GiB of 1-KiB KV objects via YCSB, we issue 1\,M reads under a
Zipf distribution over all keys with a skewness factor $\alpha$; a larger
$\alpha$ implies a more skewed distribution.  We choose $\alpha=0.9$ (the
default parameter in \cite{li21}) and $\alpha=1.2$ (for a highly skewed
workload \cite{yang16}) to study the impact of skewness.

We first consider the distribution of the number of reads to each of the SSTs
at $L_3$ in B4, in which $L_3$ is sizable enough that its SSTs may be stored
in both the SSD and the HDD.  We examine the number of SST reads at $L_3$ in
B4 in both the SSD and the HDD for $\alpha=0.9$, while the results are similar
for $\alpha=1.2$.  After the load operation, five SSTs at $L_3$ are in the
SSD, and the remaining SSTs (98 in total) are in the HDD.  Here, we present
the numbers of reads of the five SSTs in the SSD together with the top-five
frequently read SSTs in the HDD, as shown in Figure~\ref{fig:motivation}(g).
We see that there exist much more reads issued to the top-five SSTs at $L_3$
in the HDD than to the five SSTs in the SSD, implying that the read throughput
can be significantly slowed down.  The reason is that the basic schemes are
unaware of the upper-layer access patterns, and the frequently read KV objects
may be stored in the HDD.  

Figure~\ref{fig:motivation}(h) shows the percentage of read traffic that goes
to the HDD over all read traffic.  For $\alpha=0.9$, all basic schemes have
most of the read traffic (79.7-98.2\%) issued to the HDD.  For $\alpha=1.2$
(which represents a more skewed distribution), although the in-memory block
cache has absorbed much more read traffic, there are still non-negligible
fractions of HDD read traffic (35.8-49.5\%) in the basic schemes. 

Finally, we examine the actual read throughput of each basic scheme.
Figure~\ref{fig:motivation}(i) shows the read throughput of each basic scheme
for $\alpha=0.9$ and $\alpha=1.2$.  Since the basic schemes have high
fractions of HDD reads, the throughput is bottlenecked by the slow random
reads of the HDD (Table~\ref{tab:perf}).


%

\subsection{Challenges}
\label{subsec:challenge}

From \S\ref{subsec:motivation}, we summarize three challenges on designing an
efficient hybrid zoned storage system.

\paragraph{Challenge 1: Data placement.} Our goal is to put as many low-level
SSTs as possible in the SSD with high SSD utilization.  O1 shows that the
target sizes of the LSM-tree levels are inaccurate indicators for the actual
sizes of the LSM-tree levels, and O2 further shows that statically pinning
low-level SSTs for the SSD can lead to sub-optimal performance.  O3 shows that
throttling writes still cannot effectively address the limitations of basic
data placement.   Thus, a dynamic data placement algorithm is necessary for
SST storage management.  In particular, for hybrid zoned storage, it is
critical to properly place SSTs in the SSD or the HDD without compromising the
performance predictability, which is a key design feature of zoned storage
devices \cite{bjorling21}.

\paragraph{Challenge 2: Data migration.} O4 reveals that without
considering the read access patterns, a na\"ive data management scheme cannot
effectively handle the read disparity of SSTs at the same level.  Thus, the
migration of SSTs across the SSD and the HDD is inevitable (e.g., moving the
frequently read KV objects from the HDD to the SSD).  However, data migration
in hybrid zoned storage is particularly non-trivial due to the large zone
capacities (e.g., hundreds of MiB).  If the migration only moves partial data
of a zone, it causes fragmentation within the zone.  On the other hand, if the
migration moves data on a per-zone basis, it incurs substantial I/O traffic
that leads to interference to foreground activities. 




\paragraph{Challenge 3: Data caching.}  O4 also reveals that a
significant fraction of read traffic can go to the HDD, thereby degrading the
overall read performance.  Effective data placement and data migration can
reduce the amounts of HDD reads, but the frequently read KV objects may be
scattered across SSTs in the HDD.  Such frequently read KV objects should be
specifically cached for fast read performance.  Enlarging the in-memory cache
space is one possible solution but is expensive.  Caching the KV objects in the
SSD can resolve the tension of in-memory caching.  However, due to the
append-only writes of ZNS SSDs, we need to design specific caching strategies
based on the append-only interface.  Also, we should carefully balance the SSD
space, which is now shared by the WAL, low-level SSTs, and cached KV objects. 




\section{\sysname Design}
\label{sec:design}

\sysname is a middleware system that leverages {\em hints} to bridge LSM-tree
KV stores and hybrid zoned storage, with the primary goal of achieving both
high I/O throughput and low latencies in KV operations (e.g., reads, writes,
scans, and updates) subject to the limited SSD space. 

\subsection{Hints and Design Techniques}
\label{subsec:overview}

\sysname builds on three types of hints that describe the internal operations
in an LSM-tree.  The LSM-tree KV store passes a hint to \sysname along with
the corresponding operation.  Each hint is of small size with tens of bytes,
and its passing incurs limited overhead. 
\begin{itemize}[leftmargin=*] 
\item 
{\em Flushing hint}: A flushing operation passes a flushing hint that
identifies the flushed SST (at $L_0$). 
\item 
{\em Compaction hint}: A compaction operation passes a compaction hint in three
phases: (i) when the compaction operation is triggered, it passes a compaction
hint that identifies the current SSTs selected for compaction and the level
with which the SSTs are merged; (ii) when the compaction operation generates
an SST, it passes a compaction hint that specifies the level at which the SST
resides; and (iii) when the compaction operation completes, it passes a
compaction hint that identifies the SSTs generated from compaction.  
\item 
{\em Cache hint}: The in-memory block cache passes a cache hint after it
evicts a data block. The cache hint identifies the SST in which the data block
resides and the offset of the data block in the SST.
\end{itemize} 

\sysname leverages the hints to control how to manage KV objects across the
SSD and the HDD in hybrid zoned storage.  It builds on three design
techniques: 
\begin{itemize}[leftmargin=*] 
\item 
{\em Write-guided data placement.} Instead of statically reserving SSD
zones for specific LSM-tree levels as in the basic schemes
(\S\ref{subsec:motivation}), \sysname adaptively reserves SSD zones for the
low-level SSTs, by calculating how many zones are needed by each level on
the fly using both flushing hints and compaction hints.  It monitors the
actual sizes of the LSM-tree levels and avoids putting too few or too many
SSTs in the SSD. Thus, it keeps a large fraction of low-level SSTs in the
SSD and maintains high SSD utilization.  
\item 
{\em Workload-aware migration. } \sysname monitors the workload
characteristics on the fly,  so as to migrate the KV objects across the SSD
and the HDD.  Specifically, it monitors the currently occupied SSD space and
the read access patterns using both the flushing hints and
compaction hints, so as to decide when to trigger data migration operations.
It also rate-limits the data migration operations to limit their interference
to foreground activities. 
\item 
{\em Application-hinted caching. } \sysname keeps the frequently accessed HDD
data blocks that are evicted from the in-memory block cache in the SSD using
the cache hints.  Thus, it maintains high read performance for the frequently
accessed KV objects, without redundantly caching the KV objects in both the
in-memory block cache and the SSD. 
\end{itemize}

\begin{figure}[!t] 
\centering
\includegraphics[width=3in]{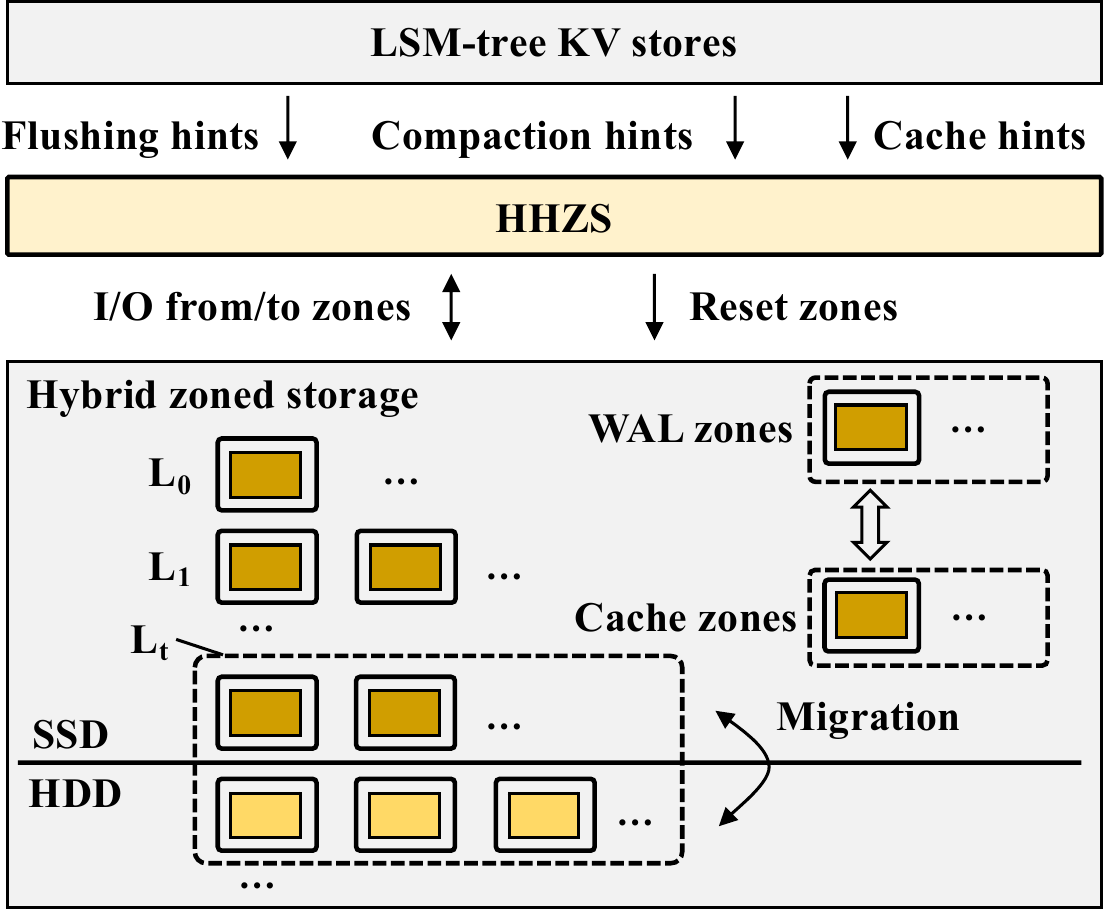}
\vspace{-3pt}
\caption{Architectural overview of \sysname.} 
\label{fig:architecture}
\vspace{-9pt}
\end{figure}

\subsection{Architectural Overview}
\label{subsec:architecture}

Figure~\ref{fig:architecture} shows the architecture of
\sysname.  An LSM-tree KV store issues I/O requests via \sysname to the zoned
storage devices. It also issues hints to \sysname, which then manages KV
objects in the zoned storage devices based on the hints. 

\paragraph{Zone organization.} With limited SSD space, \sysname dedicatedly
allocates different numbers of SSD zones to the WAL and the SSTs at different
levels.  We reserve a number of SSD zones for the WAL and the
cached data blocks, called the {\em WAL zones} and the {\em cache zones},
respectively.  We fix the total number of WAL zones and cache zones as
the pre-configured maximum size of the WAL divided by the SSD zone capacity,
while the respective numbers of WAL zones and cache zones vary based on
the access patterns (\S\ref{subsec:caching}).   This ensures that all WAL data
can be accommodated in the SSD, but some WAL zones may be switched to serve as
the cache zones if necessary.  The remaining SSD zones other than the WAL
zones and cache zones are used to keep SSTs. 

Recall that our SSD and HDD devices have zone capacities of 1,077\,MiB and
256\,MiB, respectively (\S\ref{subsec:motivation}).  Our rationale is to
configure the size of an SST to be slightly smaller than the SSD zone
capacity to achieve high space utilization in each SSD zone \cite{bjorling21},
while an SST can span across multiple HDD zones (which have smaller zone
capacities).  Thus, we set the SST size as 1,011.2\,MiB, so that it can fit in
one SSD zone (93.9\% of the SSD zone capacity) or four HDD zones (100\% of the
HDD zone capacity for the first three HDD zones, and 95\% for the last HDD
zone).  Note that \sysname does not use the write lifetime hint from RocksDB
\cite{rocksdb} for zone selection \cite{bjorling21}, as it always assigns a
new SST to one (for the SSD) or multiple (for the HDD) empty zones. 


\paragraph{Writes.}  The LSM-tree KV store generates new SSTs via either
flushing or compaction operations.  When a new SST is generated, the LSM-tree
KV store also issues a flushing hint or a compaction hint to \sysname, which 
then determines the level of the SST and hence whether the SST should be
stored in the SSD or the HDD (\S\ref{subsec:placement}).  Also, \sysname
maintains the mappings between each SST and its associated zone using the 
{\tt std::map} structure in the C++ standard template library, as in the
original ZenFS \cite{zenfs}. 

\paragraph{Reads.}  To read a KV object, the LSM-tree KV store first
checks both its MemTable and in-memory block cache for the KV object.  If it
is not found, the LSM-tree KV store requests the KV object via \sysname from
an SST in zoned storage.  \sysname first checks if the KV object resides in
the SSD (either in an SSD zone or in a cache zone). If so, it returns the KV
object; otherwise, it checks the HDD for the KV object. 


\subsection{Write-Guided Data Placement}
\label{subsec:placement}


\sysname aims to write as many low-level SSTs as possible to the SSD and
achieve high SSD space utilization based on both the flushing hints and the
compaction hints.  Given the flushing hints, \sysname attempts to write the
SSTs from flushing to the SSD, as such SSTs reside at the lowest level $L_0$.
Given the compaction hints, \sysname assigns the SSTs from compaction to
either the SSD or the HDD, depending on the levels at which the SSTs reside.
However, the actual sizes in LSM-tree levels can be significantly large (see
O1 in \S\ref{subsec:motivation}), which leads to fewer low-level SSTs that can
be actually written into the SSD.  To adapt to the varying actual sizes in
each LSM-tree level, \sysname monitors the actual sizes of the LSM-tree
levels, so as to dynamically allocate the SSD zones for the low-level SSTs
that are being written.  \sysname performs the following four steps.

\paragraph{Step~1: Calculating storage demands.} \sysname calculates the
{\em storage demand} of each LSM-tree level, defined as the number
of SSTs that need to be generated by the ongoing flushing and compaction
operations, whenever an SST is to be written.  Given the SST size and both
SSD and HDD zone capacities, the storage demand of each LSM-tree level can be
translated to the number of zones that are needed to store the SSTs for the
level in the SSD or the HDD.

\sysname first computes the storage demand of $L_0$, at which the SSTs are
generated by flushing.  The storage demand of $L_0$ is the total
size of MemTables that currently contain KV objects divided by the SST size.
Note that \sysname is unaware of the storage status of MemTables, which are
managed by the LSM-tree KV store.  Nevertheless, it knows the current number of
WAL zones being used, as each KV object in the MemTable also has a copy in the
WAL (\S\ref{subsec:lsm}). Thus, \sysname sets the storage demand of $L_0$ as
the current number of WAL zones. 

\sysname next computes the storage demand of each higher level $L_i$, where
$i\ge 1$, at which the SSTs are generated by the ongoing compaction
operations.  When a compaction operation is triggered or issues writes of
SSTs, it also sends a compaction hint to \sysname to update the storage demand.
Specifically, when there is no compaction operation that writes SSTs to $L_i$,
the storage demand of $L_i$ is zero.  When the LSM-tree KV store triggers a
compaction operation that writes SSTs to $L_i$ (we call this compaction
operation $\mathcal{C}$),
\sysname increments the storage demand of $L_i$ by the number of the selected
SSTs from $\mathcal{C}$; this is the maximum number of SSTs being generated by
$\mathcal{C}$.  Each time when $\mathcal{C}$ writes an SST to $L_i$, \sysname
decrements the storage demand by one.  When $\mathcal{C}$ finishes, \sysname
decrements the storage demand by the difference of the number of selected SSTs
minus the actual number of generated SSTs from $\mathcal{C}$. 



\paragraph{Step~2: Determining the tiering level.} Given the current
allocation of SSTs and storage demands of different LSM-tree levels, \sysname
determines the {\em tiering level} $L_t$, such that the SSTs at any level
lower or higher than the tiering level are put in the SSD and the HDD,
respectively, while the SSTs at the tiering level are stored in either the SSD
or the HDD.  Specifically, \sysname allocates the available zones in the SSD
(aside the WAL zones and cache zones) from low to high levels, until the SSTs
of a level (i.e., the tiering level) cannot be fully stored in the SSD.  Let
$A_i$ be the currently allocated zones for $L_i$, $D_i$ be the storage demand
of $L_i$, and $C_{ssd}$ be the number of available SSD zones for SSTs.  Given
that an SST is stored in one SSD zone, \sysname calculates $t = \argmin_{i}
\sum_{j=0}^{i}(A_j+D_j) \geq C_{ssd}$. 

\paragraph{Step~3: Reserving SSD zones for $L_t$.}  Given that $L_t$ is
determined, \sysname further calculates the number of SSTs at $L_t$ that can
be stored in the SSD, given by $C_{ssd} - \sum_{j=0}^{t-1} (A_j+D_j)$.  The
remaining SSTs at $L_t$ will be stored in the HDD. 

\paragraph{Step~4: Selecting zones for a written SST.}  Finally, \sysname
determines the new zone being allocated for a written SST.  \sysname attempts
to select an empty SSD zone for a written SST if (i) a flushing hint informs
that the written SST is generated from flushing; (ii) a compaction hint
informs that the written SST is from compaction and at levels lower than
$L_t$; or (iii) a compaction hint informs that the written SST is at $L_t$,
while there are still available SSD zones reserved for $L_t$.  Otherwise, if
there is no empty SSD zone or the above three conditions are violated, \sysname
selects empty HDD zones for the written SST.

\subsection{Workload-Aware Migration}
\label{subsec:migration}

\sysname performs workload-aware migration to refine the data placement in the
SSD and the HDD in the background.  We introduce two types of migration: (i)
{\em capacity migration}, which moves SSTs from the SSD to the HDD to free up
SSD zones for the low-level SSTs; and (ii) {\em popularity migration}, which
moves the frequently read SSTs in the HDD to the SSD for improved read
performance.


\paragraph{SST priorities.}  We first define the priority of an SST, so as to
determine if the SST should be migrated. The priority of an SST is based on
the SST level (indicated by both flushing hints and compaction hints) as well
as the {\em read rate} (in IOPS) of the SST, defined as the ratio between the
total number of reads and the age of the SST.  We say that for two SSTs, say
$X$ and $Y$, $X$ has a higher priority than $Y$ if (i) $X$ is at a lower level
than $Y$; or (ii) $X$ and $Y$ are at the same level while $X$ has a higher
read rate than $Y$.  Our workload-aware migration follows the priority rule to
move a higher-priority SST from the HDD to the SSD.  Note that the SSTs
selected by current compaction operations (identified by the compaction hints)
will be deleted at the end of compaction, so \sysname avoids selecting them
for migration regardless of their priorities. 


To track SST priorities, \sysname keeps the mappings between each SST and its
level, total number of reads, and age in memory.  It simply iterates all
mappings once to identify the SST priorities.  Note that the mappings incur
limited memory overhead (\S\ref{subsec:impl}). 
		
\paragraph{Capacity migration.}  When the storage demands of the lower levels
increase, \sysname needs to reserve more SSD zones for the lower levels.  Also,
it may switch the tiering level to a lower level, meaning that the SSTs in the
original tiering level should now be moved to the HDD.  Thus, \sysname performs
capacity migration to move some high-level SSTs in the SSD to the HDD.
Specifically, \sysname checks whether the tiering level has more SSTs than the
number of SSD zones reserved for the tiering level, and checks there exists
any SST in the SSD that belongs to a higher level than the tiering level.  If
either condition holds, \sysname iterates all SSTs in the SSD and selects the
SST with the lowest priority.  It moves the selected SST from the SSD to the
HDD. 


\paragraph{Popularity migration.}  Recall that some SSTs receive much more
reads than others at the same level under skewed workloads (see O4 in
\S\ref{subsec:motivation}).  \sysname performs popularity migration to move
frequently accessed SSTs in the HDD to the SSD.  Specifically, if it detects
that the read rate over all SSTs in the HDD exceeds half of the maximum random
read IOPS of the HDD (i.e., the read rate is bottlenecked by the HDD), it
starts popularity migration.  It first identifies the SST with the highest
priority in the HDD.  It then moves the SST to the SSD if the number of empty
SSD zones is larger than the total storage demands of all levels lower than
the tiering level; or swaps the SST with the one that has the lowest
priority in the SSD otherwise. 


\paragraph{Rate-limiting.} To limit the interference of migration into
foreground activities, \sysname rate-limits migration operations.  It
currently sets the rate limit as 4\,MiB/s, which is much smaller
than the HDD bandwidth (at around 200\,MiB/s).  With rate-limiting, the SST
priorities may change at the end of migration due to the migration delay and
the varying read rates of SSTs.  Nevertheless, we expect that the SST
priorities do not change drastically, while rate-limiting can effectively
mitigate the interference caused by migration. 


\subsection{Application-Hinted Caching}
\label{subsec:caching}

\begin{figure}[!t] 
\centering
\includegraphics[width=2.9in]{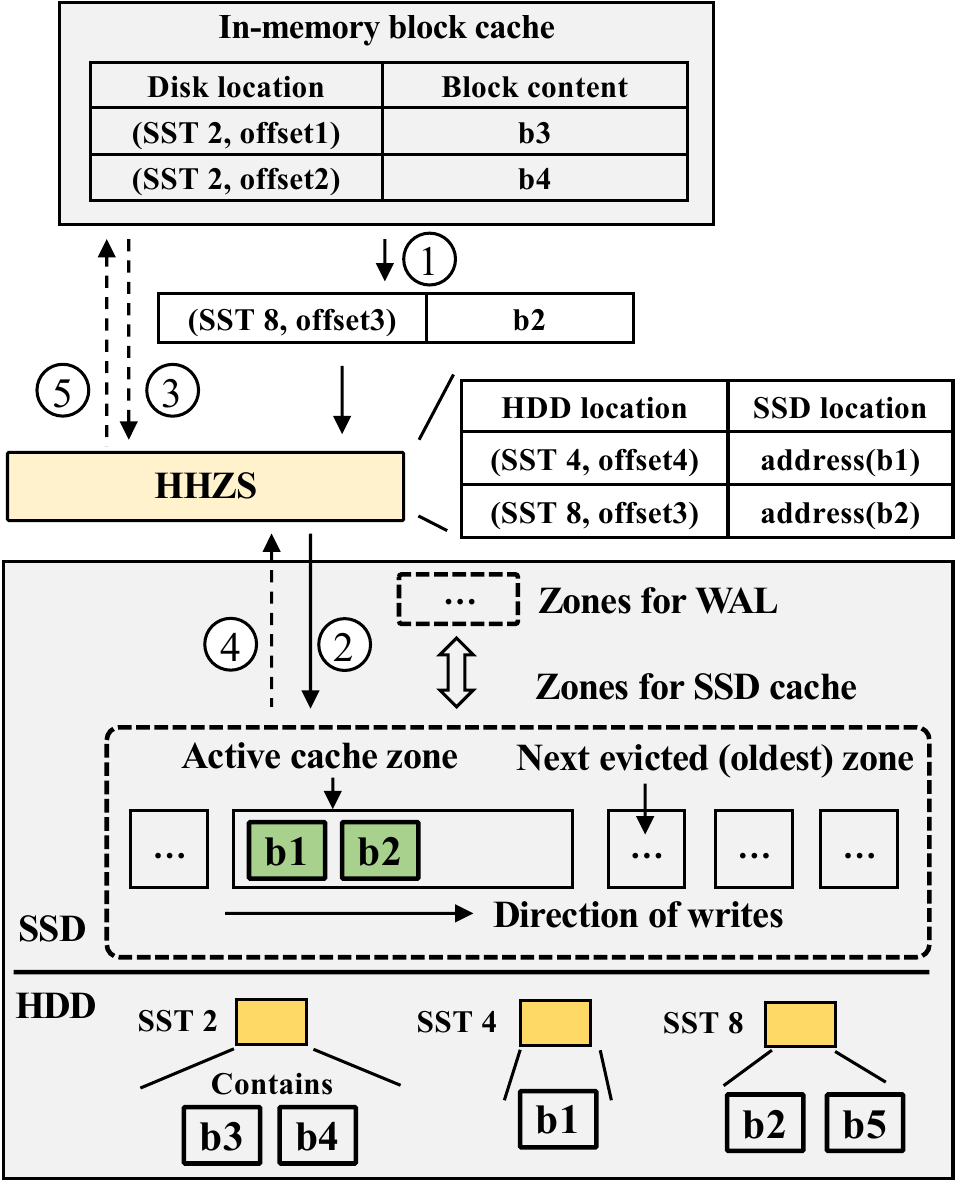}
\vspace{-3pt}
\caption{The workflow of application-hinted caching.} 
\label{fig:caching}
\end{figure}

\sysname leverages cache hints passed from the LSM-tree KV store to cache the
frequently accessed data blocks that are scattered across the HDD in the SSD.
Since the in-memory block cache in the LSM-tree KV store has already cached
the frequently read data blocks in its in-memory block cache, \sysname can
coordinate with the in-memory block cache to cache only the data blocks
evicted from the block cache to avoid redundant caching.
Figure~\ref{fig:caching} shows the workflow of application-hinted caching in
\sysname. 



\paragraph{Cache admission.}  Recall that \sysname reserves a fixed number of
WAL and cache zones (\S\ref{subsec:architecture}).  Initially, \sysname
reserves WAL zones only without any cache zone.  When \sysname receives a
cache hint (which specifies the details of an evicted data block), it converts 
an empty WAL zone into an active cache zone that stores the incoming data
blocks.  Referring to Figure~\ref{fig:caching}, the in-memory block cache of
the LSM-tree KV store indexes each cached data block by the ID of the SST at
which the data block resides and its offset in the SST. When the LSM-tree
evicts a data block (\textcircled{\raisebox{-0.9pt}{1}}), it issues a
cache hint about the evicted data block, as well as the data block content, to
\sysname.  If \sysname finds that the data block is stored in the HDD and has
not been cached in the SSD yet, it appends the data block into the active
cache zone (\textcircled{\raisebox{-0.9pt}{2}}); otherwise, it discards
the data block.  If the active cache zone is full, \sysname finds a new empty
zone (either a WAL zone or a cache zone) and converts it into a new active
cache zone. If no empty zone is available, \sysname calls cache eviction (see
below).  To efficiently identify a cached data block, \sysname maintains an
in-memory mapping table to track the HDD location (i.e., the SST ID and the
offset in the SST) for each cached data block and the SSD cache location
(i.e., the caching address in the SSD). 



\paragraph{Reads to cached data blocks.}  Reads can fetch the cached data
block if it is tracked by the in-memory mapping table.  For example, referring
to Figure~\ref{fig:caching}, suppose that the LSM-tree KV store issues a read
to data block $b1$ with the SST ID~4 and offset {\tt offset4} 
(\textcircled{\raisebox{-0.9pt}{3}}), while $b1$ is not cached by the
in-memory block cache.  \sysname finds that $b1$ is in its mapping table, so
it fetches directly $b1$ from the SSD (\textcircled{\raisebox{-0.9pt}{4}}) and
returns $b1$ to the LSM-tree KV store (\textcircled{\raisebox{-0.9pt}{5}}). 
Suppose now that the LSM-tree KV store issues a read to data block $b5$, which
does not appear in the mapping table of \sysname. Then \sysname retrieves $b5$
from the HDD.
		

\paragraph{Cache eviction.} \sysname evicts cached blocks if it runs out of
space of the reserved zones when admitting new cached blocks or writing new
WAL data. In both cases, it follows the first-in-first-out (FIFO) policy
to select an {\em evicted zone}, which is always the oldest zone among all
cache zones (Figure~\ref{fig:caching}). 
It examines all cached data blocks in the evicted zone, and removes their
corresponding HDD locations from its mapping table. It then resets the evicted
zone and makes it available for new writes. 

To identify data blocks in the evicted zone, \sysname also keeps the HDD
location and SSD cache location for each data block in an in-memory FIFO
queue (in addition to the mapping table). When \sysname admits a new data
block, \sysname also appends its HDD location and SSD cache location into the
queue.  When a zone is evicted, \sysname also dequeues the location
information of the data blocks in the evicted zone.  Thus, it can quickly
identify the HDD location and SSD cache location of each data block in the
evicted zone. 


\subsection{Implementation Details}
\label{subsec:impl}

We have built our \sysname prototype on the LSM-tree KV store RocksDB v6.22.1
\cite{rocksdb} and the C++ user-space zone-aware file system ZenFS v0.2.0
(which provides a plugin for RocksDB) \cite{zenfs}.  Specifically, we modify
RocksDB to export the three hints.  For ZenFS, we modify its zone management
module and I/O module to support multiple zoned storage devices (note that
ZenFS currently supports a single zoned storage device only) and \sysname's
design techniques by parsing the hints from the modified RocksDB.  Our
implementation includes 533\,LoC of changes in RocksDB and 2.9\,K\,LoC of
changes in ZenFS.  

\paragraph{Memory usage.} \sysname incurs limited memory overhead, which mainly
comes from the maintenance of SST priorities (\S\ref{subsec:migration}) and
cached blocks (\S\ref{subsec:caching}). 

For SST priorities, \sysname keeps the mappings between each SST and its
level, total number of reads, and age (\S\ref{subsec:migration}).  The memory
overhead is small due to the large zone capacity.  For example, if the SST ID
and other fields are of size 8~bytes each (i.e., each mapping is of size
32~bytes), for the storage of 1-TiB KV objects with a zone capacity of
256\,MiB, the memory overhead is 128\,KiB only. 

For cached blocks, \sysname maintains the in-memory mapping table for tracking
the HDD location and the SSD cache location, as well as the in-memory FIFO
queue for tracking the next evicted cached blocks (\S\ref{subsec:caching}). 
If the SST ID, the data offset in the HDD, and the SSD cache location are of
size 8~bytes each, each cached data block incurs 48~bytes for storing the
information in both the mapping table and the FIFO queue.  If the cache size
is 2\,GiB and the data block size is 4\,KiB, the maximum memory usage is
24\,MiB only.

\section{Evaluation}
 
\subsection{Experiment Settings}
\label{subsec:settings}

\noindent
{\bf Testbed.} We conduct our experiments in a machine that runs Ubuntu 21.04
LTS with Linux kernel 5.11. The machine is equipped with a 16-core Intel Xeon
Silver 4215 CPU and 96\,GiB DRAM.  It has two real zoned storage devices: a
4-TiB Western Digital Ultrastar DC ZN540 ZNS SSD \cite{zn540} and a 14-TiB
Seagate ST14000NM0007 HM-SMR HDD \cite{seagate_hmsmr}, with zone
capacities of 1,077\,MiB and 256\,MiB, respectively
(\S\ref{subsec:motivation}).  Both devices have a block size of 4\,KiB.
			 
\paragraph{Setup.}  We set the target SST size in RocksDB as 1,011.2\,MiB
(\S\ref{subsec:architecture}). We reset a zone to reclaim its space only when
the WAL data or the SST in the zone is deleted by RocksDB.  We set the
migration rate of \sysname as 4\,MiB/s. 

We configure RocksDB based on the RocksDB tuning guide \cite{rocksdb_guide}.
We set the MemTable size as 512\,MiB, flush MemTables when there are at least
two MemTables, and keep at most four MemTables in memory. We set the maximum
total size of WAL and cache zones as 2.1\,GiB (i.e., 2 SSD zones, for
application-hinted caching in \S\ref{subsec:caching}), the target sizes of both
$L_0$ and $L_1$ as 1\,GiB, and the target size of each level higher than $L_1$
as 10$\times$ the target size of its lower level.  We set a total of 12 threads
for flushing and compaction. 

To understand the performance impact of the constrained SSD size, we restrict
the available SSD size as 21.0\,GiB (i.e., 20 SSD zones) and do not limit the
HDD size, while we also evaluate the impact of different SSD sizes (Exp\#5).
The total KV object size is 200\,GiB, and each KV object has a 24-byte key and
a 1,000-byte value.  The in-memory block cache is 8\,MiB (the default in
RocksDB).  

Before evaluating each workload in all experiments, we always first clear the
storage and load the KV objects, so that each workload is independently
evaluated.  We enable direct I/O for SST reads and writes, so as to reduce the
impact of the page cache. 

\paragraph{Schemes.}  We consider the basic schemes B1-B4
(\S\ref{subsec:motivation}) and the automated placement (referred to as 
{\em AUTO}) in the hybrid LSM-tree KV store SpanDB \cite{chen21}.  AUTO
controls the maximum level, such that all LSM-tree levels up to the maximum
level are put in fast storage (i.e., the SSD in our case) based on the current
throughput and the remaining space in fast storage.  We re-implement AUTO in
our testbed based on the open-source implementation of SpanDB.  Specifically,
when the SSD throughput is less than 40\% or higher than 65\% of the
sequential write throughput (Table~\ref{tab:perf}),  AUTO increases or
decreases the maximum level by one, respectively.  When the remaining SSD
space is less than 13.3\% of the total SSD space, AUTO fixes the maximum level
as one; if the remaining space of SSD is less than 8\% of the total SSD space,
AUTO does not write any SST data to the SSD.  Note that AUTO reserves the SSD
space for the WAL, as in \sysname. 

\subsection{Results}
\label{subsec:results}

\noindent
{\bf Summary of findings.}  \sysname achieves the highest throughput compared
with all
baselines under different YCSB workloads (Exp\#1).  Our breakdown analysis
shows that each design technique contributes to the performance gain of
\sysname (Exp\#2).  Also, \sysname maintains its high throughput subject to
different skewness factors (Exp\#3), read-write ratios (Exp\#4), and SSD
space sizes (Exp\#5).  Its default migration rate of 4\,MiB/s incurs low
interference (Exp\#6). 
	

\paragraph{Exp\#1 (YCSB workloads).} We first evaluate \sysname with the YCSB
benchmark \cite{cooper10}.  YCSB provides six core workloads: A (50\% reads
and 50\% updates), B (95\% reads and 5\% updates), C (100\% reads), D (95\%
latest reads and 5\% writes), E (95\% range queries and 5\% writes),
and F (50\% reads and 50\% read-modify-writes). Each workload has 1\,M
operations.  All workloads (except workload~D) follow a Zipf distribution with
a skewness parameter $\alpha=0.9$ \cite{li21}, while workload~D reads the
latest written keys.  For workload~E, the starting point of each scan is
selected according to the Zipf distribution.  We compare \sysname with B3 (the
fastest basic scheme; see \S\ref{subsec:motivation}) and AUTO. 

\begin{figure}[!t] 
\centering
\begin{tabular}{@{\ }c}
\includegraphics[width=3.3in]{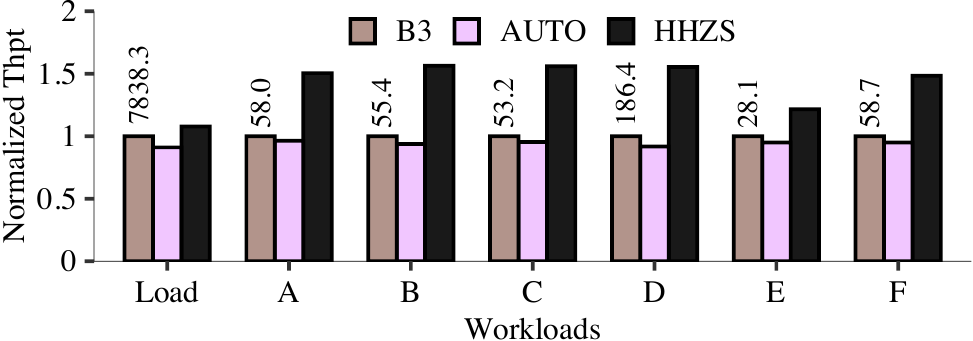} 
\vspace{-3pt}\\
\parbox[t]{3.2in}{\small (a) Normalized throughput with respect to B3 (the
number above each bar is the throughput of B3 in OPS)}  
\vspace{9pt}\\
\includegraphics[width=3.3in]{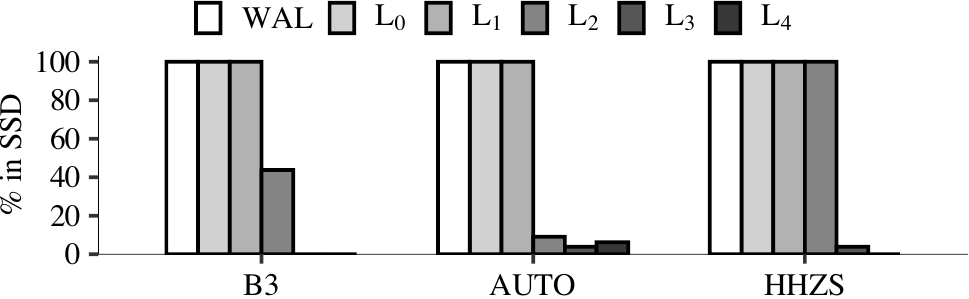}
\vspace{-3pt}\\
{\small (b) Percentages of data in the SSD at the end of workload A}
\end{tabular}
\vspace{-6pt}
\caption{Exp\#1 (YCSB workloads).}
\label{fig:exp1}
\vspace{-10pt}
\end{figure}

Figure~\ref{fig:exp1}(a) shows the throughput of loading all KV objects and
workloads A-F, and Figure~\ref{fig:exp1}(b) shows the percentages of data that
resides in the SSD at each level at the end of workload A (similar results are
also observed in workloads B-F).  \sysname shows higher load
throughput than B3 and AUTO (with a gain of 7.8\% and 18.3\%,
respectively). In particular, it shows higher throughput in workloads~A-F than
B3 and AUTO (with a gain of 21.0-56.4\% and 28.0-69.3\%,
respectively), mainly because it manages to store all SSTs at $L_0$-$L_2$ and
some frequently read SSTs in $L_3$ in the SSD (Figure~\ref{fig:exp1}(b)).  AUTO
has even slower throughput than B3 in all cases, as its maximum level tuning
can be sensitive to the throughput thresholds that are used to update the
maximum level.  This causes AUTO to put frequently accessed SSTs in the HDD.
Compared with B3, \sysname better utilizes SSD zones with migration and
caching, while compared with AUTO, \sysname correctly puts low-level SSTs in
the SSD with write-guided data placement.


\paragraph{Exp\#2 (Performance breakdown).} We provide a performance breakdown
to show how each design technique contributes to the performance gain of
\sysname.  We start with B3 and then consider the following schemes: 
%
\begin{itemize}[leftmargin=*] 
\item {\em B3+M:} It combines B3 with workload-aware migration (M)
(\S\ref{subsec:migration}).  It moves the SSTs at $L_0$-$L_2$ in the HDD to
the SSD under $M$. Note that it does not move SSTs at $L_3$ from the HDD to
the SSD (even though there are empty zones in the SSD), as B3 requires
to keep all SSTs at $L_3$ and $L_4$ in the HDD.
\item {\em P:} It deploys write-guided data placement (P), without 
migration and caching. 
\item {\em P+M:} It deploys write-guided data placement (P) and workload-aware
migration (M), without caching. 
\item {\em P+M+C:} It deploys write-guided data placement (P), workload-aware
migration (M), and application-aware caching (C).  It is equivalent to the
whole design of \sysname. 
\end{itemize}

\begin{figure}[!t] 
\centering
\includegraphics[width=3.3in]{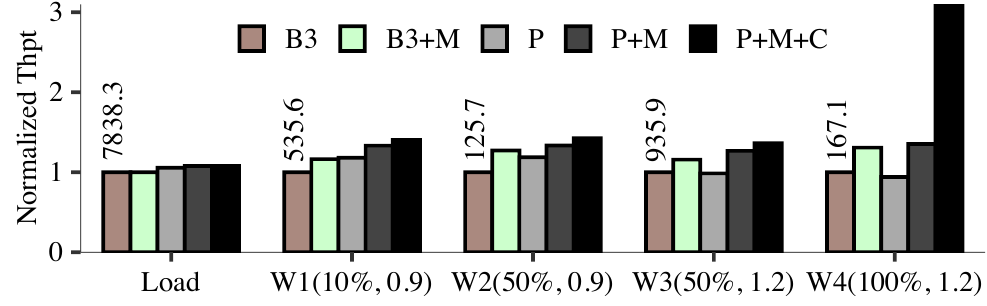}
\vspace{-3pt}
\caption{Exp\#2 (Breakdown). The number above each bar is the throughput of
B3 in OPS.  The x-axis shows the load and the percentages of reads and skewness
factors for W1-W4.}
\label{fig:exp2_breakdown}
\vspace{-6pt}
\end{figure}

We first load 200\,GiB KV objects in each scheme.  After loading the KV
objects, we run a workload of 5\,M operations following a Zipf key
distribution for a read-write ratio and a skewness factor $\alpha$.  We
consider four types of workloads: W1 (10\% reads and $\alpha=0.9$); W2 (50\%
reads and $\alpha=0.9$); W3 (50\% reads and $\alpha=1.2$); and W4 (100\% reads
and $\alpha=1.2$). We choose $\alpha=1.2$ for high skewness, given that
practical workloads have skewness factors no more than 1.2 \cite{yang16}.  

Figure~\ref{fig:exp2_breakdown} shows the normalized throughput of different
schemes with respect to B3.  Compared with B3, write-guided data placement (P)
itself shows improved throughput in some cases, but is sometimes worse.  P has
higher throughput by 5.5\%, 18.0\%, and 18.8\% in load, W1, and W2,
respectively, but less throughput by 1.4\% and 5.9\% in W3 and W4,
respectively.  For load, the throughput improvement of P comes from the higher
fractions of SSTs at $L_0$ and $L_1$ being stored in the SSD. For W1 and W2,
B3 may write the SSTs at $L_0$ and $L_1$ to the HDD due to the amplified
actual sizes of $L_2$, while P always writes the SSTs at $L_0$ and $L_1$ to
the SSD.  For W3 and W4, both P and B3 issue non-negligible fractions of reads
to the SSTs at $L_2$ being stored in the HDD, so P cannot achieve performance
gains and it has even less throughput than B3. 


With migration, both B3+M and P+M improve their counterparts without migration
(B3 and P, respectively) in W1-W4, and migration works better with P than B3.
As depicted by Figure~\ref{fig:exp2_breakdown}, P+M has higher throughput than
B3+M by 14.5\%, 8.8\%, 9.4\%, and 5.9\% in W1, W2, W3, and W4,
respectively.  The reason is that both P and M aim to allocate SSD zones
adaptively based on the storage demands.  As a result, P+M can efficiently fix
the placement violation of P under workload changes.  In contrast, B3 follows
static SSD space reservation and does not use the available SSD zones to store
the frequently accessed SSTs. 


Caching further improves the throughput of P+M in W1-W4, and the improvement
is more obvious for a higher fraction of reads or a higher skewness factor.
For load, caching has no effect and the throughput remains the same as P+M.
P+M+C increases the throughput of P+M by 5.3\%, 6.8\%, 7.4\%, and
173.7\% in W1, W2, W3, and W4, respectively.  A higher fraction of reads
implies that more copies in the SSD cache are less likely updated, so the SSD
cache receives more cache hits.  Also, a higher skewness factor implies that
the SSD cache has more read hits as more reads are aggregated on a smaller set
of KV objects.  


\paragraph{Exp\#3 (Impact of the workload skewness). } We study the impact of
the workload skewness (i.e., the skewness factor $\alpha$) on the performance of
\sysname.  We run workloads with 5\,M operations and vary $\alpha$ from 0.8 to
1.2. We also fix 50\% of reads and 50\% of writes. Figure~\ref{fig:exp3_skew}
shows the throughput of B3, AUTO, and \sysname.  \sysname has
27.3-43.3\% of throughput gain for different workload skewness
over B3. It also has 51.6-77.1\% of throughput gain over AUTO. The
throughput gain comes from the higher SSD utilization (compared with B3) or the
placement of more low-level SSTs in the SSD (compared with AUTO). 

\begin{figure}[t]
\begin{tabular}{@{\ }cc}
\parbox[t]{1.6in}{
  \includegraphics[width=1.58in]{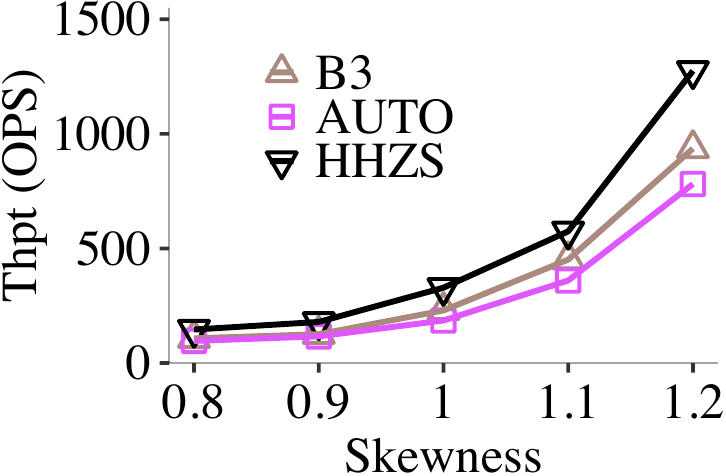} 
  \vspace{-15pt}
  \caption{Exp\#3 (Impact of the workload skewness).}
  \label{fig:exp3_skew}
}
&
\parbox[t]{1.6in}{
  \includegraphics[width=1.58in]{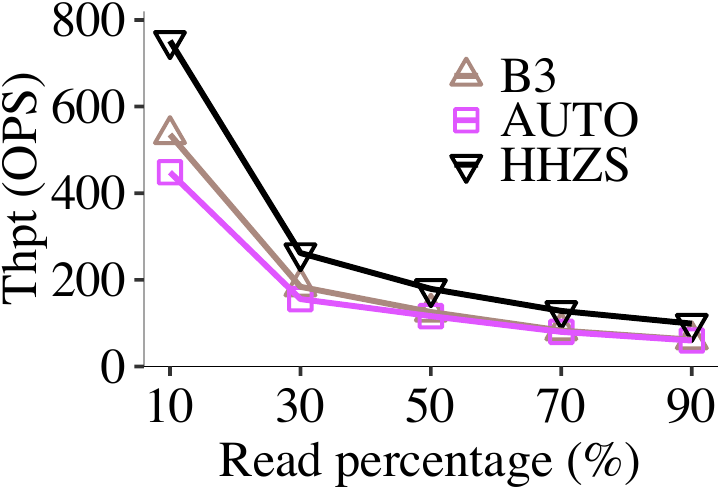} 
  \vspace{-15pt}
  \caption{Exp\#4 (Impact of read-write ratios). }
  \label{fig:exp4_rw}
}
\end{tabular}
\vspace{-6pt}
\end{figure}

\paragraph{Exp\#4 (Impact of the read-write ratio). } We study the impact of
the read-write ratio. We run a workload with 5\,M operations.  We vary the
percentage of reads from 10\% to 90\%, and fix $\alpha=0.9$.
Figure~\ref{fig:exp4_rw} shows the throughput of B3, AUTO, and \sysname.  For
small read percentages of 10\% and 30\%, \sysname has 40.4\% and
42.4\% higher throughput than B3, respectively, as well as 67.7\%
and 68.4\% higher throughput than AUTO, respectively.  For high read
percentages at least 50\%, \sysname has 42.5-60.0\% and
54.1-65.3\% higher throughput than B3 and AUTO, respectively.  Due to the slow
HDD random reads, smaller read percentages imply higher throughput since a read
is much slower than a write.  Thus, the overall performance of different
schemes is mainly constrained by the slow HDD reads, while \sysname accelerates
reads by performing popularity migration and caching. 

\begin{figure}[!t] 
\centering
\begin{tabular}{@{\ }c}
\includegraphics[width=3.3in]{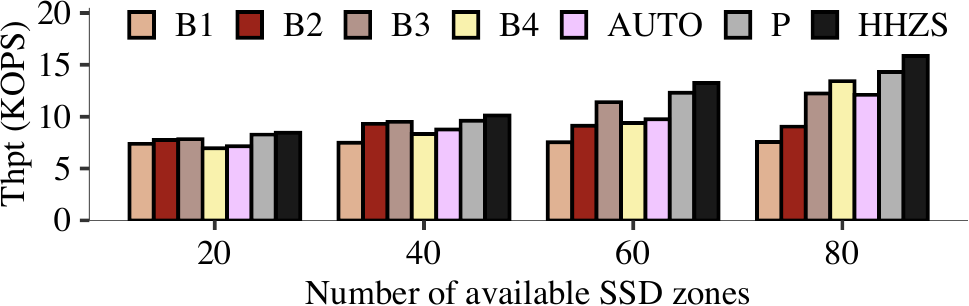} 
\vspace{-3pt}\\ 
{\small (a) Throughput of loading KV objects}  
\vspace{3pt}\\ 
\includegraphics[width=3.3in]{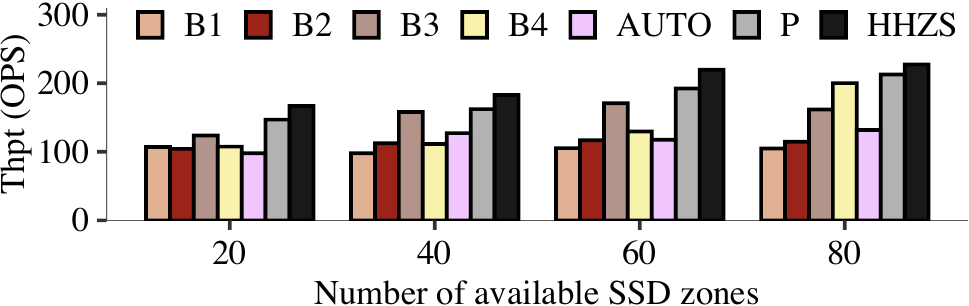} 
\vspace{-3pt}\\ 
{\small (b) Throughput with 50\% reads and $\alpha=0.9$}
\end{tabular}
\vspace{-10pt}
\caption{Exp\#5 (Impact of the SSD size).}
\label{fig:exp5}
\vspace{-10pt}
\end{figure}

\paragraph{Exp\#5 (Impact of the SSD size).} We examine how the SSD size
affects the performance of \sysname.  We vary the size of the available SSD
space from 21.0\,GiB (i.e., 20~zones) to 84.1\,GiB (i.e., 80~zones).
We first load 200\,GiB KV objects and compare
\sysname with the four basic schemes B1-B4, AUTO, and P (i.e., write-guided
data placement in Exp\#2).  Figure~\ref{fig:exp5}(a) shows the throughput of
different schemes in load.  P is robust in maintaining high throughput in
various SSD sizes.  Compared with the basic schemes and AUTO, its throughput
is 5.5\%, 1.1\%, 7.9\%, and 6.6\% higher than the best schemes among the basic
schemes and AUTO under 20, 40, 60, and 80 available SSD zones, respectively.
The full design \sysname further increases the load throughput of P by 2.2\%,
5.3\%, 7.7\%, and 10.8\% in four SSD sizes, respectively, with the help of
capacity migration. 

We also run a workload of 1\,M operations with 50\% reads, 50\% writes, and
$\alpha=0.9$ for different SSD sizes. Figure~\ref{fig:exp5}(b) shows the
results.  
For all SSD sizes, both P and \sysname have higher throughput than other
schemes. Under four SSD sizes, compared with the best among all basic schemes
and AUTO, P increases the throughput by 18.8\%, 2.6\%, 12.5\%, and 6.2\%,
respectively, while \sysname increases the throughput by 34.8\%, 15.5\%,
28.5\%, and 13.6\%, respectively.

\paragraph{Exp\#6 (Impact of the migration rate).} Recall that \sysname
rate-limits migration to reduce interference with foreground traffic 
(\S\ref{subsec:migration}).  We study the impact of the migration rates on
the tail latencies. We do not consider application-hinted caching and use P+M
as in Exp\#3. We vary the migration rates from 1\,MiB/s to 64\,MiB/s. We load
200\,GiB of 1-KiB KV objects, and run a workload with 50\% reads, 50\% writes,
and $\alpha=0.9$.  We focus on the read latencies. 

Figure~\ref{fig:exp6_migthpt} shows the 99th, 99.9th, and 99.99th percentile
read latencies versus the migration rate.  The 99th percentile latency remains
similar for all migration rates, with less than 10\% of variations. The 99.9th
percentile latency is the lowest for migration rates of 2\,MiB/s and 4\,MiB/s,
while the highest 99.9th percentile latency is on the migration rate of
64\,MiB/s, which is 23.0\% and 25.4\% higher than those of 2\,MiB/s and
4\,MiB/s, respectively. It shows that a large migration rate increases the
interference with foreground reads. The 99.99th percentile latency shows an
increasing trend with the increasing migration rate, and increases by 104\%
from 1\,MiB/s to 64\,MiB/s.  Our default setting selects 4\,MiB/s as the
migration rate for our evaluation without significantly increasing the tail
latency. 

\begin{figure}[!t]
\centering
\includegraphics[width=2in]{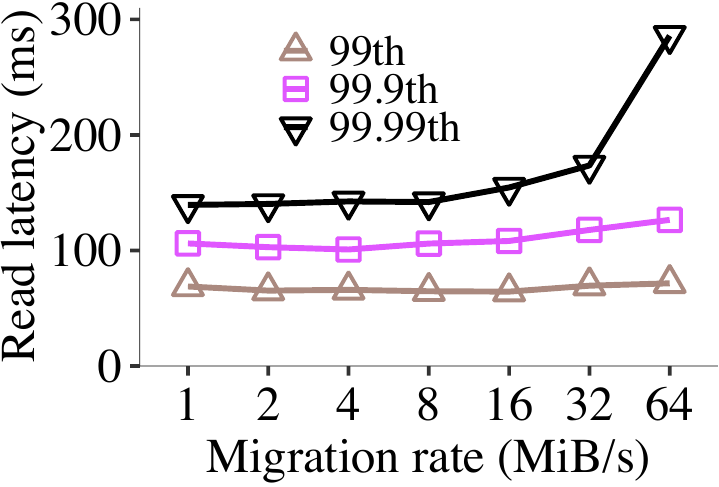}
\vspace{-9pt}
\caption{Exp\#6 (Impact of the migration rate).}
\label{fig:exp6_migthpt}
\vspace{-6pt}
\end{figure}



\section{Related Work}
\label{sec:related}

\noindent
{\bf Zoned storage.}  Prior to zoned storage, previous studies expose internal
SSD features to applications (e.g., Software-Defined Flash \cite{ouyang14},
Application-Managed Flash \cite{lee16}, and Open-Channel SSDs
\cite{bjorling17, huang17}) for fine-grained storage management. Such
proposals are specific to SSDs.  In contrast, zoned storage provides a unified
zoned interface for both ZNS SSDs and HM-SMR HDDs (\S\ref{subsec:zoned}). 

There have been various storage system proposals for SMR
drives \cite{jin14, pitchumani15, manzanares16, wu16, yao17, macko17, xie18,
xie19, yao19, yao19sealdb, yang18, aghayev20} and ZNS SSDs \cite{choi20, han21,
bjorling21, maheshwari21}.  To name a few, 
Duchy \cite{xie18} proposes a hybrid storage design of SSDs and SMR drives by
enabling SSD caching and regulating the number of written zones in SMR drives
to bound the recovery time.  GearDB \cite{yao19} is an LSM-tree KV store for
HM-SMR HDDs, using a gear compaction algorithm to eliminate on-device
garbage collection. 
BlueStore \cite{aghayev20} adapts RocksDB to HM-SMR HDDs and provides a
zoned storage backend for distributed storage.   
Choi et al. \cite{choi20} design a garbage
collection scheme based on hot-cold data segregation for ZNS SSDs. ZNS+
\cite{han21} proposes a new interface to accelerate segment compaction in
zone-aware log-structured file systems. Bj{\o}rling et al.  \cite{bjorling21}
adapt F2FS \cite{lee15} and RocksDB to ZNS SSDs, and show improved throughput
and decreased tail latencies. Maheshwari \cite{maheshwari21} proposes an
abstraction for storing variable-size data in ZNS SSDs. Compared with existing
zoned storage designs, \sysname proposes a hinted design for hybrid zoned
storage. 



\paragraph{KV stores for hybrid storage.}  Some KV stores combine
traditional NAND-flash-based SSDs with the faster but more expensive
non-volatile memory (NVM) for cost-efficient and high-speed storage
\cite{yoon18, kaiyrakhmet19, li20, raina20, chen21}.  GTSSL \cite{spillane11}
statically pins an LSM-tree level to the SSD or the HDD, and re-inserts
frequently read KV objects into low levels to reduce HDD reads.  Mutant
\cite{yoon18} dynamically balances the cost-performance trade-off by organizing
the placement of SSTs among fast and slow storage. SLM-DB
\cite{kaiyrakhmet19} uses the B+-tree in NVM for indexing and stores KV
objects in a single-level organization in SSDs.  HiLSM \cite{li20} proposes
log-structured designs in NVM and efficient data migration from NVM to
SSDs.  PrismDB \cite{raina20} monitors the access frequencies of KV objects, and
places high-frequency KV objects in NVM and low-frequency KV objects in SSDs.
SpanDB \cite{chen21} combines NVMe SSDs and SATA SSDs, by storing most KV
objects in SATA SSDs and relocating write-ahead logs and the top levels of the
LSM-tree to NVMe SSDs.  Some KV stores (e.g., HiKV \cite{xia17}, Flashield
\cite{eisenman19}, and AC-Key \cite{wu20}) use DRAM for efficient indexing or
caching, and use SSDs for persistent storage. 
In contrast, \sysname supports LSM-tree KV stores on hybrid storage with
ZNS SSDs and HM-SMR HDDs. 

\paragraph{Hinted storage.}  Application hints improve storage performance in
many aspects \cite{patterson95, sarkar96, liu09, ouyang14, kumar20, tai21,
macedo22}.  Informed prefetching and caching \cite{patterson95} uses
application hints to exploit I/O parallelism and dynamically allocate file
buffers in file systems. 
CLient-Informed Caching (CLIC) \cite{liu09} uses client-based hints to improve
read hit ratios of storage server caches.
Quiver \cite{kumar20} proposes an informed storage cache policy for deep
learning training jobs in GPU clusters. PAIO \cite{macedo22} uses the context
of each I/O request for configuring I/O optimization policies.  Hints are also
used in boosting the performance of distributed systems \cite{sarkar96,
adams21}, flash memories \cite{ouyang14}, and NVMe devices \cite{tai21}.  In
contrast, \sysname focuses on the co-design of the LSM-tree (i.e., RocksDB)
and the zoned file system (i.e., ZenFS), and leverages hints provided by the
LSM-tree KV store to manage data in hybrid zoned storage.

\section{Conclusion}

\sysname is a middleware system that leverages hints issued by the upper-layer
LSM-tree KV store to perform efficient data management on hybrid SSD and HDD
zoned storage. It builds on three techniques to aim for high performance: (i)
write-guided data placement adaptively reserves SSD zones for low-level SSTs;
(ii) workload-aware migration dynamically moves SSTs between the SSD and the
HDD with rate-limiting; and (iii) application-hinted caching keeps copies of
the frequently accessed data blocks in the SSD.  Experimental results
demonstrate the throughput gain of \sysname prototype over the baselines in
hybrid zoned storage.

\bibliographystyle{abbrv}
\bibliography{reference}

\end{document}